\documentclass{PoS}

\title{News from Cosmic Ray Air Showers \\(Cosmic Ray Indirect - CRI Rapporteur)}

\ShortTitle{Rapporteur: Cosmic Ray Indirect}

\author{\speaker{Frank G. Schr\"oder}\\
        Bartol Research Institute, Dept.~of Phys.~and Astr., Univ.~of Delaware, Newark, DE, U.S.A.;\\
Institute of Nuclear Physics, Karlsruhe Institute of Technology (KIT), Karlsruhe, Germany\\
        E-mail: \email{fgs@udel.edu}}

\abstract{This rapport summarizes the cosmic-ray indirect (CRI) session of the 36\textsuperscript{th} ICRC conference. 
Updated measurements from several air-shower arrays with higher precision were shown leading to the discovery of new features in the energy spectrum: 
HAWC measures a softening of the light component (p+He) around $10^{13.5}\,$eV; measurements of the Pierre Auger Observatory show that the second knee is a smooth feature extending at least over the range of $100-200\,$PeV and that the energy spectrum between the ankle and the cut-off cannot be described by a simple broken power law.
Measurements of the mass composition confirm that the composition is a varying mixture of protons and nuclei at least up to several $10\,$EeV. 
Hadronic interaction models constitute a significant uncertainty in the interpretation of measurements, but a joint effort of several collaborations helps to better assess their deficiencies, e.g., by quantifying the muon deficit in the models over the shower energy.
Still, general trends in the average mass composition over energy are consistent for all state-of-the-art models. 
Anisotropy measurements with higher precision generally confirm earlier results, too.
A change of the amplitude and phase of the equatorial dipole provides another indication that in the energy range between the second knee and the ankle there likely is a transition from Galactic to extragalactic sources.
However, neither the most energetic Galatic nor the extragalactic sources have been discovered, yet, which remains a primary science goal of the field. 
Next to more exposure, an increase of measurement accuracy and decrease of systematic uncertainties will provide future progress. 
Therefore, it is particularly exciting that new experiments are built and existing experiments upgraded to increase the accuracy for the measurement of the energy and mass composition, e.g., by combining radio antennas with particle detectors. 
Last but not least, there is a trend that experiments are designed such that they can target cosmic rays, photons, and neutrinos at the same time, which will facilitate multi-messenger astrophysics at the highest energies.
}

\FullConference{36th International Cosmic Ray Conference -ICRC2019-\\
		July 24th - August 1st, 2019\\
		Madison, WI, U.S.A.}

\begin{document}

\section{Introduction}
The 36\textsuperscript{th} ICRC demonstrated manifold progress regarding the most energetic Galactic and extragalactic cosmic rays that are usually measured indirectly by cosmic-ray air showers.
More than 100 talks and more than 150 posters were presented in the 'cosmic-ray indirect' (CRI) sessions of this conference complemented by several plenary talks.
Hence, it is impossible to provide a complete rapport of the numerous material presented at the conference, and I had to perform a difficult personal selection from the large number of contributions. 
For this, I relied mostly on the content presented directly at the conference, i.e, the posters displayed and the talks presented, and only to some extent on the written material in the proceedings. 
In any case, the references given here are to the corresponding proceedings, and I will go through them sorted by topics. 

In addition to the scientific results themselves, I want to highlight three particular ways of how the work was done because these build the foundation of some of the most important discoveries presented at this ICRC.
First, an increase in quality results from higher experimental precision and more thorough studies of systematic uncertainties. 
We have surpassed the era of speculative interpretation of poor data, and important experimental results are now based on increasingly well understood instruments and solid statistical interpretations. 
Second, the large international collaborations operating the major observatories for air-shower detection do not only compete, but also collaborate with each other. 
By this joint effort, they achieve a higher sky coverage and a more robust study of systematic effects, increasing the physics impact beyond the potential of any of the individual experiments.
Third, theoretical models used for the interpretation of the measurements get thoroughly tested against the manifold cosmic-ray measurements, and in some case also against multi-messenger data. 
These tests are a cumbersome effort requiring detailed simulations studies, but it is the only way to test whether a model is consistent with the complete available experimental data.
Last but not least, we have seen several efforts for future experiments, often upgrades aiming at an increase of accuracy, which shows that cosmic-ray science is a living field. 

The solutions to the big questions about the origin of the highest energy cosmic rays produced inside and outside of our Galaxy are still ahead.
Nonetheless, a number of important steps were taken, resulting in a consolidation of the general picture that emerged over the past years:
At all investigated primary energies, i.e., at least up to several $10^{19}\,$eV, cosmic rays are composed of a mixture of atomic nuclei - where the composition varies with energy.
Cosmic rays at energies below the second knee around $10^{17}\,$eV seem to be of Galactic origin.
There is a transition from Galactic to extragalactic origin in the energy range between the second knee and the ankle, i.e. roughly around $10^{18}\,$eV.
At least the extragalactic cosmic rays at highest energy (above $10^{19}\,$eV) seem to originate mainly from Galaxies in our cosmic neighborhood (i.e., from closer than $100-200\,$Mpc).

Although the interpretation of measurements suffers from several shortcomings regarding models of hadronic interactions in air showers, acceleration mechanisms, and (inter)galactic magnetic fields, to my understanding, the above findings are relatively robust. 
Nevertheless, many questions remain open, in particular, there may be several source populations contributing with different mass compositions at different energies to the cosmic-ray flux, but none of the suggested source candidates above several $10^{15}\,$eV has been experimentally confirmed.
Since we do not know by what measurements or type of observations these sources will be revealed, progress in all subtopics of our field remains important.

\begin{figure}[t]
    \centering
    \includegraphics[width=0.99\linewidth]{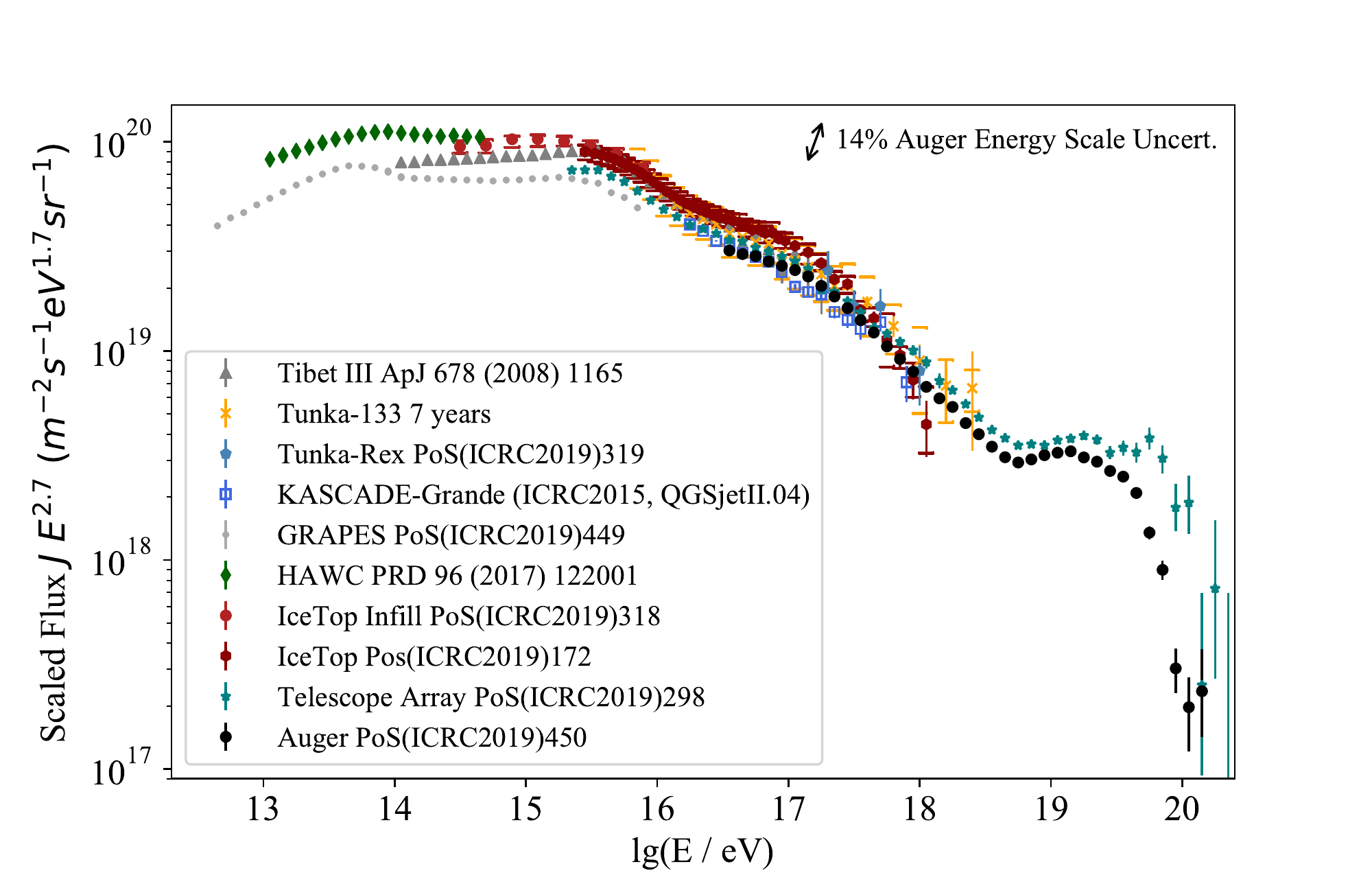}
    \caption{Energy spectrum of high-energy cosmic rays obtained from air-shower measurements \cite{TibetApj2008, PoS319, Tunka133_7years, KG_ICRC2015, PoS449, HAWC_PRD2017, PoS318, PoS172, PoS298, PoS450}. 
    Different measurement techniques are used by the experiments, and systematic uncertainties have been investigated in varying detail. 
     The effect of an uncertainty of the absolute energy scale is shown exemplary for the Pierre Auger Observatory \cite{PoS231}. 
}
    \label{fig_spectrum}
\end{figure}

\section{Energy Spectrum}
Many collaborations provided new \cite{PoS319, PoS449, PoS318} or updated measurements on the cosmic-ray energy spectrum \cite{PoS172, PoS298, PoS450}. 
In several cases, the energy range was extended towards lower energies by dedicated analysis methods, and the quality of the measurement improved, e.g., by accumulating additional statistics and by a thorough study of systematic uncertainties. 
Naturally, the spectra have different quality, reaching from a simple proof-of-principle that  an experiment works as expected, to hybrid measurements featuring low systematic uncertainties. 
In particular, hybrid measurements using fluorescence telescope have the advantage that the absolute energy scale relies on external calibration measurements and features minimal dependence on hadronic interaction models. 
In future, also radio measurements may provide an independent calibration of the absolute scale \cite{TunkaRexLOPESenergyScale, AERAenergyScale}.

Figure \ref{fig_spectrum} shows energy spectra presented in the cosmic-ray indirect session and a selection of spectra published earlier since not all experiments provided updates at this ICRC. 
Generally, the spectra are in agreement with each other when taking into account statistical, systematic, and scale uncertainties. 
Only at the very highest energies above the cut-off, there is some tension between the flux measured by Telescope Array and the Pierre Auger Observatory that can only partly be explained by the observation of different parts of the sky \cite{PoS234}. 
It also remains open to what extent the cut-off is due to the GZK effect, i.e., energy-loss due to interactions of protons and nuclei with CMB photons during their propagation, or due to the maximum acceleration energy of the sources. 
Nevertheless, at lower energies, the results of both experiments are compatible. 

The increase in precision and accuracy leads to several new insights:
\begin{itemize}
\item HAWC measured a softening in the spectrum of light particles (p+He) around $10^{13.5}\,$eV \cite{PoS176}. 
Some of the relatively new space experiments orbiting the Earth are expected to provide direct measurements in this energy region soon \cite{PoS032} and may investigate this feature separately for protons and He nuclei.

\item Auger measured the second knee with remarkable accuracy and showed that the second knee is not a sharp feature, but a softening that extends over at least a factor of two in energy from about $100-200\,$PeV \cite{PoS225}. 
This extended range might explain why earlier measurements of the second knee position varied in energy by more than the stated uncertainties of various experiments.

\item The increased precision of Auger also revealed that the spectrum above the ankle is more complex than the simple broken power law that was used until now to describe the cut-off. 
It is no longer obvious whether a power-law is an appropriate description at all the cosmic-ray energy spectrum above the ankle. 
At least another feature has to be included in the description between the ankle and the cut-off \cite{PoS450}.
\end{itemize}

\section{Mass Composition}
For air-shower measurements, the mass composition of cosmic rays is traditionally estimated from either the distribution of the atmospheric depth of the shower maximum, $X_\mathrm{max}$, or measurements of the muon number relative to the size of the electromagnetic shower component.
$X_\mathrm{max}$ measurements at the highest energies ($E \gtrsim 10^{17}\,$eV) mostly rely on the well-studied air-fluorescence technique.
It provides highest accuracy when combined with surface detectors for air-shower particles. This type of hybrid detection is the primary method used by the Telescope Array (TA) \cite{PoS013} and the Pierre Auger Observatory \cite{PoS004}.
At lower energies, hybrid measurements of $X_\mathrm{max}$ are yet less common, and mostly particle measurements are used for estimating the mass composition.

\begin{figure}[p]
    \centering
    \includegraphics[width=0.99\linewidth]{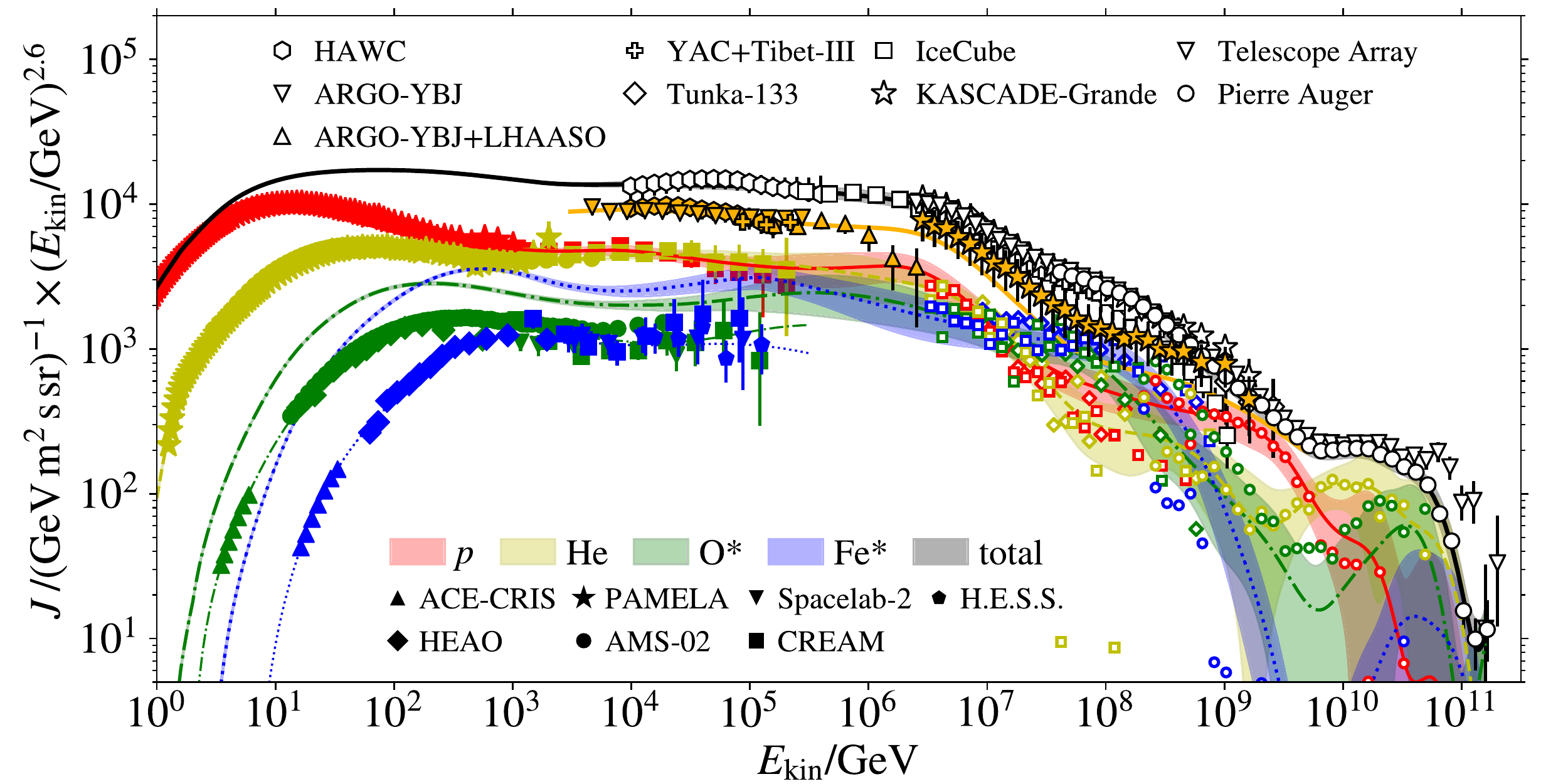}
    ~~~~\\
    ~~~~\\
    \vspace{-0.5cm}
    \includegraphics[width=0.99\linewidth]{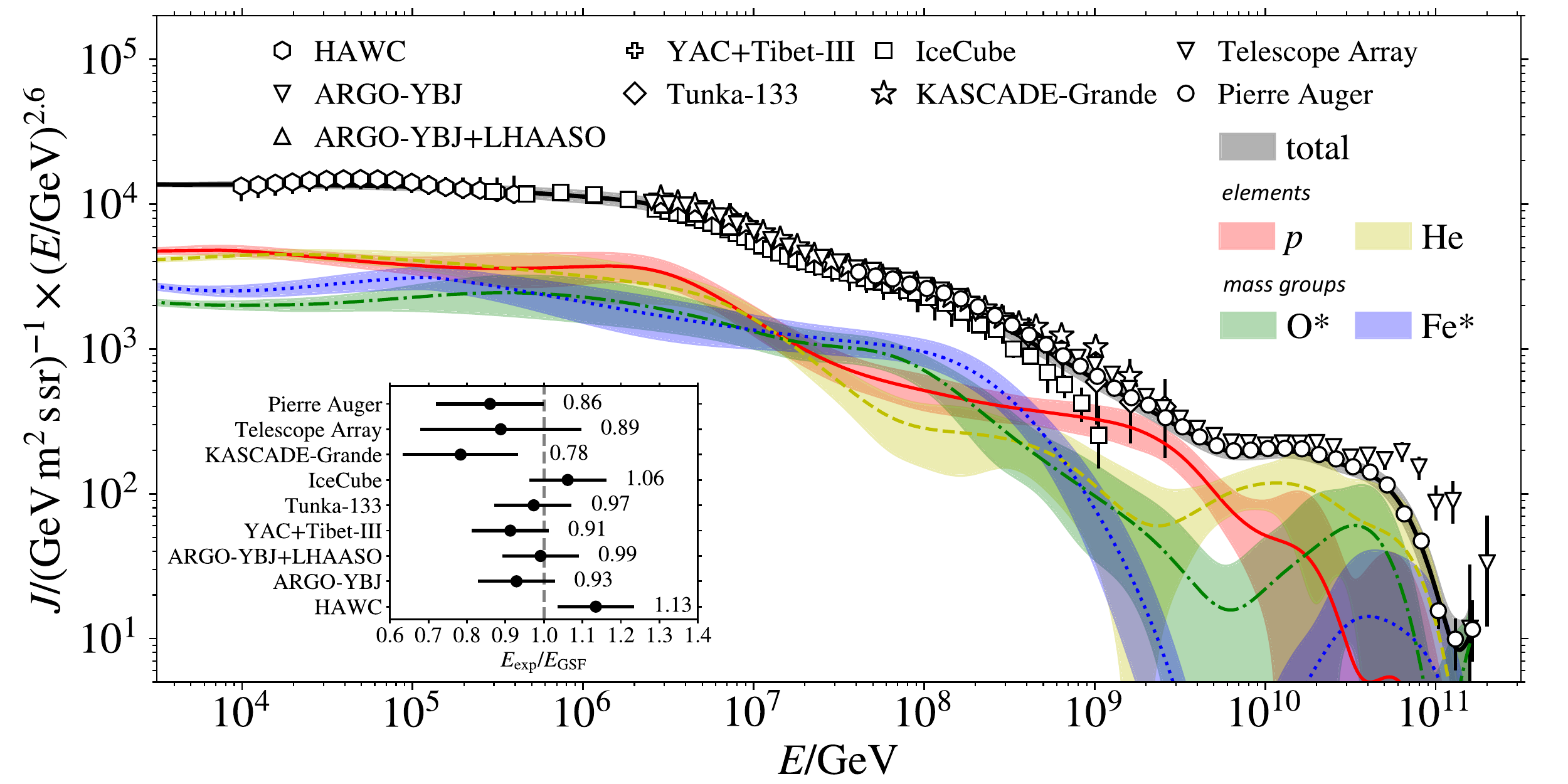}
        \caption{Combined fit of the cosmic-ray energy spectra and its mass composition measured by various experiments. 
        While taking into account the experimental uncertainties, the fit allows in addition for a constant shift of the energy scale of each experiment (inlet in the lower figure). 
        This shift of the energy scales aims to bring the measured spectra in agreement, and generally is within the scale uncertainties.
        The method of this Global Spline Fit (GSF) is presented in Ref.~\cite{GSF_ICRC2017} including references to the used data sets. 
        Hans Dembinski was so kind to update his fit including several new measurements presented at the CRI session of this ICRC, in particular: the HAWC p+He spectrum \cite{PoS176} and the mass fractions measured by IceCube \cite{PoS172}, as well as most of the new all-particle spectra presented in Fig.~\ref{fig_spectrum} (new data of direct measurements presented at this ICRC have not yet been included). 
        The individual data points and the resulting fit are shown in the top figure, a zoom to the CRI energy range is presented in the bottom figure.
}
    \label{fig_GSF}
\end{figure}

For the highest energy Galactic cosmic rays, i.e., at about $10^{15} - 10^{18}\,$eV, IceCube presented the result of an analysis combining the measurements of high-energy muons in the ice with surface measurements of the same air showers by IceTop, where the IceTop signal is typically dominated by electromagnetic particles \cite{PoS172, PoS394}.
The information of these two detectors is combined by a neural network reconstructing the fractions of four mass groups over energy and the all-particle spectrum. 
These results are included in Figs.~\ref{fig_spectrum} and \ref{fig_GSF}.
In the same energy range, KASCADE-Grande presented an updated analysis using the latest hadronic interaction model Sybill 2.3c, confirming earlier results of a softening of the heavy component and a hardening of the light component around the second knee \cite{PoS306}. 

\begin{figure}[t]
    \centering
    \includegraphics[width=0.46\linewidth]{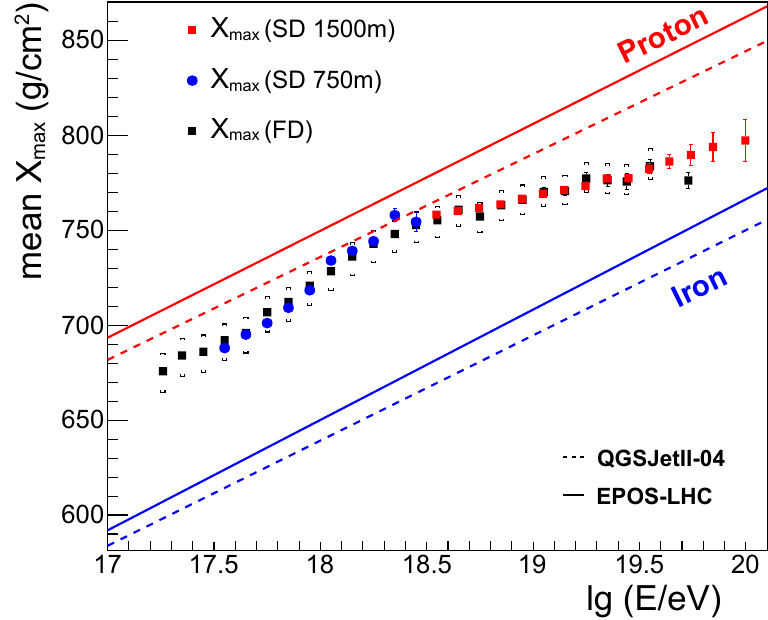}
    \hfill
    \includegraphics[width=0.52\linewidth]{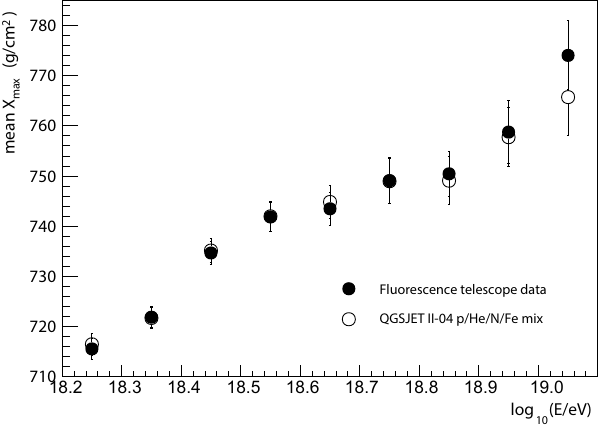}
    
    \caption{Mean $X_\mathrm{max}$ measured by the Pierre Auger Observatory (left; adapted from \cite{PoS440}) and by the Telescope Array (right; adapted from \cite{PoS280}). 
    The Auger measurements are without significant selection bias and can be compared directly to predictions generated by Monte-Carlo simulations using different hadronic interaction models. 
    For TA, the detection bias is included in the simulations (see text for mass fractions used in this plot).
}
    \label{fig_Xmax}
\end{figure}

For the highest energies, TA \cite{PoS280} and Auger \cite{PoS482, PoS440} presented updates based on increased statistics and improvements in the analysis. 
Due to its larger aperture and longer operation time, Auger accumulated more statistics which pays off in higher sensitivity, especially at energies above $10^{19}\,$eV.
Already at the last ICRC, it had been shown that there is no contradiction in the $X_\mathrm{max}$ measurements between both experiments \cite{PoS(ICRC2017)522}.
Although the latest update of this joint analysis was performed for the UHECR 2018 conference \cite{UHECR2018_AugerTA_Xmax}, the conclusion is generally supported by new results presented at this ICRC.

While Auger sees evidence for a fraction of heavier elements than He around the ankle \cite{PoS482}, a light mass composition cannot be excluded by TA. 
Nonetheless, also for TA, a mixed mass composition of approx.~$57\,\%$ p, $18\,\%$ He, $17\,\%$ N, and $8\,\%$ Fe provides the best fit to the measured $X_\mathrm{max}$ distribution around the ankle between $10^{18.2}$ and $10^{19.1}\,$eV (using the QGSJETII.04 hadronic interaction model; numbers yet without uncertainties). 
Auger measurements extend to higher energies: The average mass composition becomes heavier with energy above $10^{18.5}\,$eV.
Furthermore, the decreasing size of $\sigma(X_\mathrm{max})$ indicates that the composition above $10^{18.5}\,$eV becomes less mixed and can only contain a small fraction of light elements around $10^{19.5}\,$eV \cite{PoS482}. 

Furthermore, a trend of changing composition over energy can be assessed in a relatively model-independent way by the elongation rate (slope of the mean $X_\mathrm{max}$ over energy) \cite{PoS482}.
Auger clearly confirms a break around the ankle reported earlier, i.e., the composition is lightest around the ankle and becomes heavier towards the second knee at lower energies as well as towards the cut-off at higher energies (Fig.~\ref{fig_Xmax}). 
TA measurements are compatible with a break in the elongation rate, but cannot exclude a constant elongation rate within present statistical uncertainties. 

\subsection{Radio measurements}
Radio measurements of $X_\mathrm{max}$ have become an alternative to the established air-Cherenkov and air-fluorescence measurements \cite{SchroederReview2017, HuegeReview2016}, and several radio experiments presented respective updates \cite{PoS319, PoS205, PoS246}.
In particular, LOFAR demonstrated that systematic uncertainties for the radio $X_\mathrm{max}$ measurement of an individual air shower are well studied and not larger than for the established techniques. 
However, for translating individual $X_\mathrm{max}$ measurements into a measurement of the mass composition over energy, additional systematic uncertainties come into play regarding the energy scale, the selection efficiency, and the aperture. 
Uncertainties for the energy scale are lowered to the same level as for the standard techniques \cite{PoS362}, but the remaining uncertainties for the efficiency and aperture are difficult to estimate. 
Due to the interplay of two emission mechanisms (geomagnetic and Askaryan), this is more complicated for radio arrays than for particle detectors.
Nevertheless, also for the calculation of the exposure of radio arrays progress has been demonstrated \cite{PoS331}, which would facilitate the operation of radio arrays as a stand-alone technique. 
For antennas as complement to hybrid detectors, maturity has already been demonstrated and antennas will be part of several planned detector upgrades (see Sec.~\ref{sec_futureProjects}).

\subsection{Global Spline Fit}
At the last ICRC, a global spline fit (GSF) was presented to combine results from different experiments using the uncertainties stated by the respective collaborations. 
For this proceeding, the GSF fit was updated using the all-particle energy spectra presented by IceCube, TA, and Auger, the p+He spectra by HAWC and Tibet, and the mass fraction by IceCube (Fig.~\ref{fig_GSF}). 

The GSF assumes a constant shift of the energy scale between experiments, and the spectra are shifted correspondingly, taking into account their uncertainties. 
Although a possible energy dependence of the different scales is neglected and the scale shift of an experiment by the GSF does not necessarily imply a deviation from the unknown true energy scale, the GSF scale is better than just a relative, arbitrary scale.
Since it is based on the statistical propagation of experimental uncertainties, the more accurate direct measurements at low energies have a strong influence on the scale, and the GSF enables to link the regimes of direct and indirect measurements.
As seen in the cosmic-ray direct session of the ICRC, there are also new measurements of space experiments, some of them yet preliminary. 
In the next few years, these will create a significant overlap in the TeV to PeV range to indirect measurements by HAWC, Tibet, TAIGA, GRAPES, and LHAASO, which will further constrain the combined fit.

%Add some anisotropy figures (HAWC+IceCube with multipoles; Auger dipole strengths over energy).
\section{Anisotropy}
Galactic cosmic rays arrive at Earth almost isotropically.
Only weak anisotropy amplitudes of the order of a permill or lower (depending on energy) have been observed. 
In particular, a statistically significant dipole anisotropy has been observed at almost all energies, except for the transition region between a few PeV and EeV where $5\,\sigma$ statistical significance has not yet been achieved \cite{AhlersAnisotropyReview2017, PoS306}.
Stronger anisotropies of amplitudes of several percents have only been observed for extragalactic cosmic rays at the highest energies \cite{AugerScience2017}.
Regarding anisotropy measurements, several updates were presented at this ICRC, confirming earlier findings with higher precision.

\begin{figure}[t]
    \centering
    \includegraphics[width=0.99\linewidth]{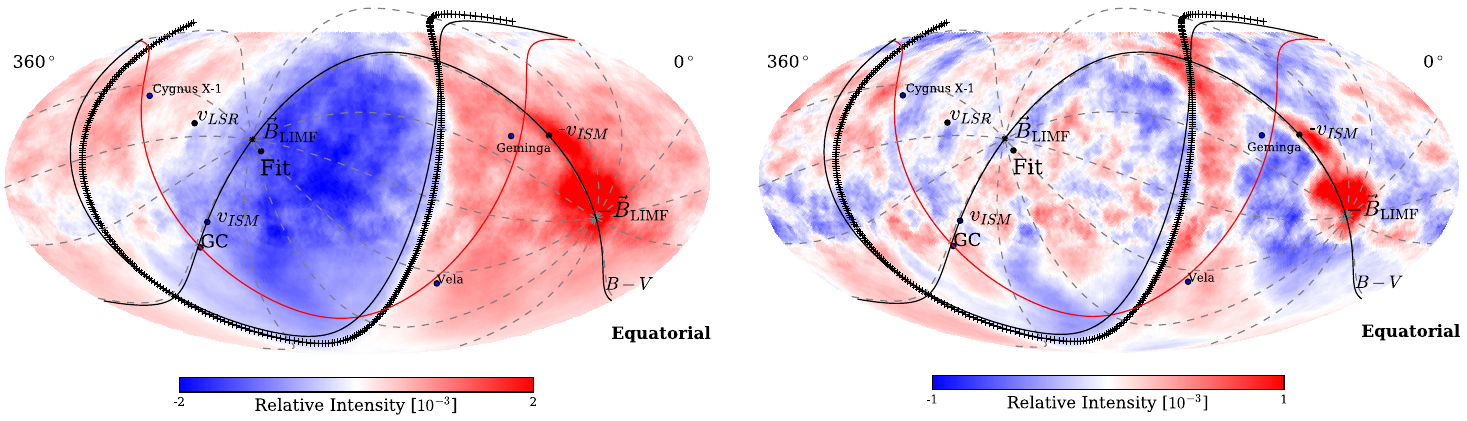}
        \caption{Anisotropy measured by HAWC and IceCube at $10\,$TeV before (left) and after (right) subtracting a multipole fit (Fig.~from \cite{PoS014}; originally published in \cite{HAWC_IceCube_anisotropy} - see there for a detailed explanation).
}
    \label{fig_HAWC_IceCubeAnisotropy}
\end{figure}

A combined analysis by HAWC and IceCube at $10\,$TeV was able to determine several multipole moments of a flux map covering almost the entire sky, which corresponds to anisotropies at different angular scales (see Fig.~\ref{fig_HAWC_IceCubeAnisotropy}) \cite{PoS014}.
Anisotropies observed by Tibet \cite{PoS488} and LHAASO \cite{PoS263} confirm the general picture. 
It was suggested that the dipole component of the anisotropy in the TeV to PeV range might be explained by the Compton Getting effect, i.e., the movement of the solar system in the Milky Way.
This hypothesis needs to be studied with more accurate measurements of the energy evolution of the anisotropy \cite{PoS488}.

\begin{figure}[t]
    \centering
    \includegraphics[width=0.99\linewidth]{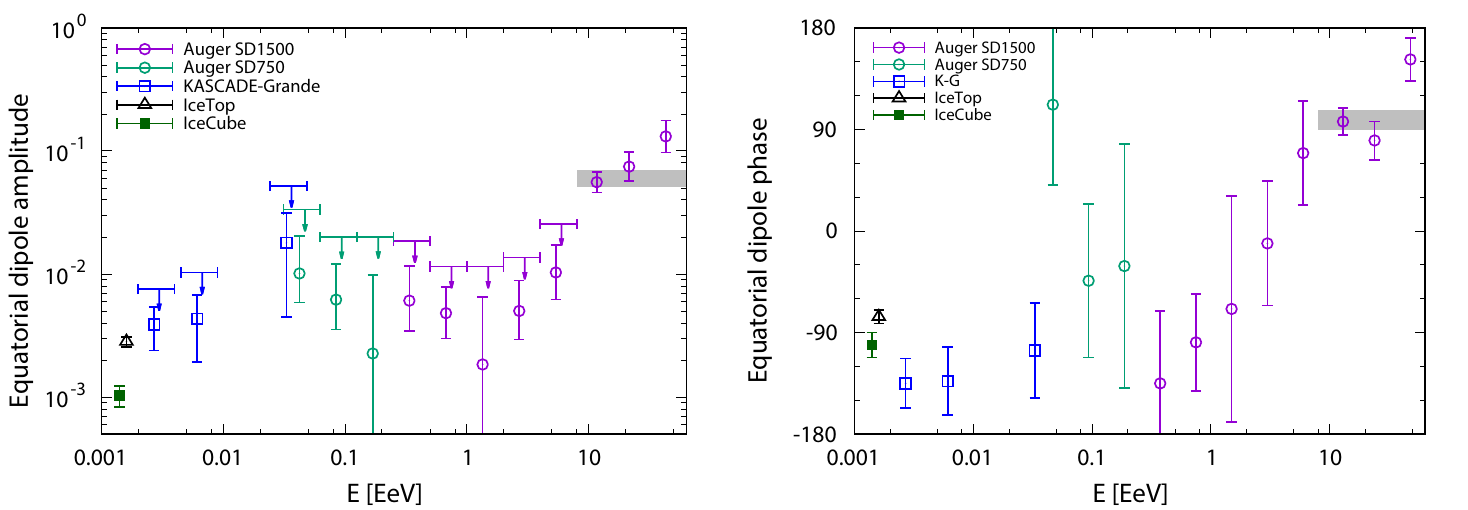}
        \caption{Amplitude (left) and phase (right) of the equatorial dipole of the large-scale anisotropy at ultra-high energies (from \cite{PoS408}).
}
    \label{fig_dipole}
\end{figure}

At the highest energies, the Pierre Auger Collaboration reports that the strength of the dipole anisotropy found earlier increases with energy \cite{PoS408}, and reaches a strength of the equatorial dipole of more than $10\,\%$ for energies above approximately $10^{19.5}\,$eV (Fig.~\ref{fig_dipole}).
The evolution of the dipole phase with energy is one of the indications for a transition from Galactic to extragalactic sources around $1\,$EeV \cite{PoS206}.
Since many data analyses of cosmic rays are based on the assumption of an approximately  isotropic flux, the relatively large size of the anisotropy demands for a check of systematic uncertainties or biases implied by that assumption. 
A step in this direction was done by a simulation study showing that the flux (and its energy spectrum) measured at Earth is expected to agree within a few percents with the extragalactic flux impinging our Galaxy \cite{PoS468}.

Another indication for the extragalactic origin of the most energetic cosmic rays is the correlation of their arrival directions with the structure of the nearby universe ($d \lesssim 100\,$Mpc). 
The arrival directions of the highest energy ($E \gtrsim 40\,$EeV) cosmic rays measured by Auger were found to be correlated with the positions of galaxies of various types from several catalogs. 
The strongest correlation is with starburst galaxies, but the interpretation is not trivial because all galaxy types are correlated with the large-scale structure of the local universe and some galaxies may belong to several types and undergo different stages.
In addition to statistical searches, there is the chance of directly discovering a source by observing a hotspot. 
Nevertheless, for the two candidates, i.e., the TA hotspot visible from the Northern hemisphere \cite{PoS310}, and the Auger hotspot around Cen A visible from the Southern hemisphere \cite{PoS439}, the statistical significance has not yet reached the discovery level. 

In summary, anisotropy studies are one of the most powerful tools to search for the yet unknown origin of the most energetic cosmic rays. 
Better models of the Galactic and intergalactic magnetic fields will help to improve propagation models required for a detailed understanding. 
On the experimental side, the next major step will be the resolution of the all-particle anisotropy into mass groups.
Several collaborations are undertaking upgrades of their detectors to provide the necessary accuracy for this endeavor (see Sec.~\ref{sec_futureProjects}).

\section{Origin}
The fundamental quest of the origin of cosmic rays contains several questions:
What are their sources? By what mechanisms are cosmic rays accelerated? How do they propagate to Earth?
These questions have not been solved, but  progress has been made on several aspects. 
In contrast to other areas of our field that are dominated by the large experimental collaborations, this progress is mostly provided by individuals or small collaborations.
Therefore, it is difficult to give a complete and balanced overview. 
Thus, I refrain from listing all possible source candidates and only mention a small personal selection of few contributions presented in the CRI section.

Shockfront acceleration, e.g., in supernova remnants, remains a plausible scenario for the dominant acceleration mechanism. Improved hybrid simulations of relativistic ions provide a plausible explanation, why spectra of supernova shockfronts can be steeper than a power-law index of $\gamma= -2$ \cite{PoS209}. 
In another study, it was shown that helium nuclei might be accelerated more efficiently than protons or other nuclei \cite{PoS228}.
However, the relation to the different power-law indices of the p and He spectra observed at Earth was not discussed and will need to take into account also propagation effects.
Realistic simulations of propagation effects require accurate models of Galactic and extragalactic magnetic fields, which remain another topic of active research.
For extragalactic cosmic rays, the possibility of a magnetically guided propagation along galaxy filaments in the large-scale structure was presented. 
Due to the propagation along filaments, the TA hotspot could be caused by cosmic rays accelerated in the Virgo cluster \cite{PoS315}.
Generally, such a guided propagation over large angular distances can have interesting consequences on the interpretation of small-scale anisotropies, such as correlation studies with possible source catalogs.

Generally, scenarios for the origin of ultra-high-energy cosmic rays need to be tested against experimental data. 
Ideally, all high-quality data available are taken into account, such as the cosmic-ray mass composition, energy spectrum, and sky maps, as well as observational data or limits of other messengers. 
An example of such a multi-messenger test including neutrino measurements by IceCube and cosmic-ray measurements by Auger is presented in Ref.~\cite{PoS364}. 
The result suggests that ultra-high-energy cosmic rays may be comprised of two components: The dominant component would feature a heavy composition above $10\,$EeV, but a sub-dominant protonic component could reach out to much higher energies.

Another scenario for extragalactic sources was presented in Ref.~\cite{PoS196}, and tested by a combined fit against various measurements of the Pierre Auger Observatory. 
The specific scenario is for Gamma-Ray Bursts (GRBs) as sources and takes into account how the interaction of the accelerated cosmic rays within the GRB shells affects the spectral slope of the protons and nuclei escaping.
Therefore, the source spectra as well as propagation effects of extragalactic cosmic rays contribute to the ankle feature in the all-particle energy spectrum observed at the Earth. 
While the original dip-model explaining the ankle by the energy loss of protons due to pair creation neglects that there is a significant fraction of heavier nuclei, also those nuclei undergo energy losses during their propagation \cite{UngerPRD2015}. 
In the scenario of Ref.~\cite{PoS196}, fitting the cosmic-ray mass composition requires a fraction of Galactic cosmic rays being present until the energy region of the ankle.
Consequently, also the Galactic-to-extragalactic transition would contribute to the ankle feature, which implies that the most energetic Galactic cosmic rays have an energy of a few EeV.

Regarding the origin of Galactic cosmic rays, the progress concentrates on the region of lower energies (below the knee; mostly GeV and TeV) accessible by direct cosmic-ray and gamma-ray observations \cite{PoS031}.
Based on gamma-ray observations, the Galactic Center remains a candidate for the acceleration of protons at least to the PeV region \cite{HESS_GC_Nature2016}, but it is not yet clear how the origin of the most energetic Galactic cosmic rays around and beyond the second knee can be explained.

\section{Hadronic Interactions in Air Showers}
Hadronic interaction models remain one of the major systematic uncertainties in the interpretation of air-shower measurements. 
Apart from that, understanding the hadronic and particle physics in the air showers is science goal by itself and an area of traditional cooperation with the particle-physics community.
Thus, there is hope that ongoing and planned measurements at accelerators, such as proton-oxygen collisions at LHC \cite{PoS235}, will provide the basis for improvements in hadronic interaction models - finally resulting in a higher overall accuracy of indirect cosmic-ray measurements.
Since a concise overview of hadronic interactions in air showers is given a separate highlight proceeding \cite{PoS005}, I will only summarize a few selected contributions on this topic here.

\begin{figure}[t]
    \centering
    \includegraphics[width=0.99\linewidth]{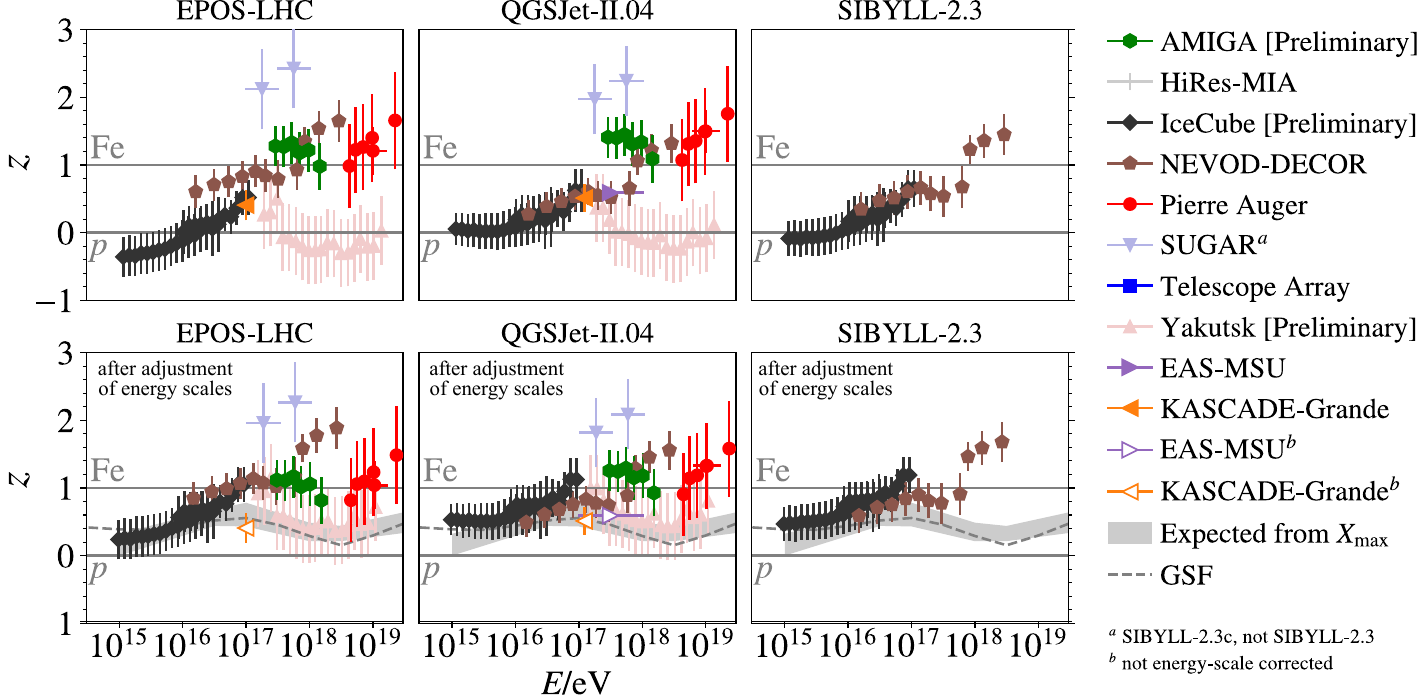}
    \caption{Muon measurements over energy by various experiments compared to the predictions of hadronic interaction models before and after adjustment of the energy scales by the GSF fit (Figure adapted from Ref.~\cite{PoS214}). 
}
    \label{fig_MuonMeasurements}
\end{figure}

Another example of experimental collaborations increasing their impact by cooperation is a combined analysis of muon measurements by air-shower arrays over many order of magnitude in energy \cite{PoS214}.
Refined analyses of existing measurements, as well as new measurements, were included, such as the pure measurement of the muon density by the AMIGA underground 
detectors at the Pierre Auger Observatory \cite{PoS411}.
The joint analysis of several experiments confirms that all state-of-the-art hadronic interaction models predict too few muons (Fig.~\ref{fig_MuonMeasurements}). 
This muon problem seems to grow with energy and becomes significant around $100\,$PeV (likely it already starts at lower energies): The measured muon numbers exceed the predicted ones by several $10\,\%$ (but not more than a factor of 2), where the exact number depends on the energy and on the hadronic model used for comparison.
However, the relative shower-to-shower fluctuations of the muon numbers seem to be in agreement with hadronic interaction models \cite{PoS404}.
Nonetheless, there are other observables not reproduced by the models, e.g., the evolution of the muon number with zenith angle \cite{PoS177}. 
Even in the TeV energy range, well below the center-of-mass energy of contemporaneous accelerator experiments, there still are significant differences between hadronic interaction models of the order of $10-20\,\%$ \cite{PoS351, PoS417}. 

The electromagnetic component seems to be understood significantly better than the muonic one, but the quantitative level of accuracy of the models is not clear. 
At least for the position of the shower maximum, there is a small tension between simulations and measurements \cite{PoS482}.

In addition to the essential progress in dedicated accelerator experiments targeting the phase space relevant for air showers \cite{PoS446, PoS207, PoS349, PoS188, PoS387}, there are at least two further ways how progress can be achieved.
In some cases, it is possible to obtain information on the hadronic interactions from the air-shower measurements themselves. 
One example is the seasonal variations of muons measured with extreme precision by IceCube. 
The non-linearity of the correlation with the effective atmospheric temperature is linked to the pion-to-kaon ratio, which itself depends on the hadronic interactions in the shower \cite{PoS894}.
A second example was presented in a simulation study indicating that a precise measurement of the probability distribution of the muon number at a given energy can provide information on neutral pions in air showers \cite{PoS226}.

Another way of progress will be to continue the development of Monte Carlo simulations based on hadronic interaction models and test them for consistency with experimental data of all kinds and in the complete relevant energy range. 
Such developments will be facilitated by a major upgrade of the main working horse for such simulations studies:
The CORSIKA simulation package is being transferred to C++ and undergoes several significant improvements \cite{PoS236, PoS181, PoS399}. 
The new CORSIKA 8 will have a modular structure enabling the whole community to contribute to the future development of this software.

\section{Other Topics}
While the essence of our field is fundamental research in astroparticle physics at the highest energies, there are several merits to other areas.

One of the practical applications of cosmic rays is muon tomography \cite{PoS377}.
Several groups pursue research in this area, e.g., to monitor the status of volcanoes \cite{PoS381, PoS275} or to detect movements of ancient buildings \cite{PoS201}. 
Another application is the use of cosmic-ray detectors for atmospheric physics, either directly by observing the atmosphere with the cosmic-ray detectors, or indirectly by observing changes in the detected cosmic-ray signals due to atmospheric events.
One of the examples for the latter category is the study of atmospheric electric fields by their influence on the radio emission of air showers \cite{PoS416}.
There are many more examples, as summarized in a dedicated overview contribution \cite{PoS018}.

Outreach is another way to increase the impact of our research field beyond its core science goals. 
Many collaborations make an effort to reach out to school students and the general public, e.g., by organizing outreach events, on-site tours, or public webpages.
Furthermore, there are a few dedicated outreach projects, such as CREDO for distributed cosmic-ray detection, which uses, on the one hand, smartphone sensors and, on the other hand, small scintillation detectors at public locations such as schools \cite{PoS428}. 
Due to the known situation of the latter, the data are easier to interpret, in this way increasing the potential for scientific outcomes besides fulfilling the outreach purpose. 

Open data becomes increasingly important not just for outreach purposes, but also for the leverage of scientific research. 
Generally, experimental collaborations have different policies what experimental data to share and how to share them, but in most cases, at least a part of the data is shared either by open websites or upon request. 
KASCADE set an example a few years ago by making its measurements available via a website for scientific analysis (an approach very common in the astronomy community). 
This effort is now extended towards the Tunka experiment \cite{PoS284}, and also data of other experiments have been included in this KASCADE data center (KCDC) \cite{KCDC_EPJC}. 
At this ICRC, we have seen an example that opening the KASCADE data to the community indeed motivates further research \cite{PoS453}.
Consequently, open data can indeed enhance the scientific merit of an experiment.

%\section{Experiments}
%
%
%\subsection{Ongoing experiments}
%
%HAWC, GRAPES, Tibet-ASgamma, TAIGA, IceCube, Pierre Auger Observatory, Telescope Array, LHAASO started operation.
%
%Radio arrays:
%AERA, LOFAR, Tunka-Rex, CODALEMA

\section{New Projects and Upgrades} \label{sec_futureProjects}
One of the most exciting news at this ICRC was about the numerous new projects and upgrades of existing air-shower arrays that are forthcoming in our field.
They will increase the precision and, most importantly, the total measurement accuracy over a wide range of energies. 
A review of the manifold science goals is beyond the scope of this rapport. 
Nevertheless, the recent Astro2020 decadal survey provides a good overview of the general science goals targeted by indirect cosmic-ray measurements (and much beyond), see, e.g., Refs.~\cite{Astro2020_GCR, Astro2020_UHECR}. 
Again, it is hard to track all individual experimental upgrades, but the following list summarizes some of the more advanced plans for upgrades and new detectors presented at the conference.

\begin{itemize}
\item LHAASO in China: LHAASO is a multi-detector experiment under construction.
It features a dense $1.3\,$km$^2$ air-shower array of electromagnetic and muon detectors complemented by an extremely dense $78,000\,$m$^2$ water-Cherenkov detector array in the center \cite{PoS217, PoS219}. 
When completed, it will target Galactic cosmic rays over many orders of magnitude in energy up to about $10^{18}\,$eV. 
The combination of complementary detector systems is expected to make this the most accurate air-shower array in the Northern hemisphere. 
Very promising results were presented at this ICRC using parts of LHAASO already completed. 
Examples are the measurement of the lateral distribution of muons \cite{PoS472} and the observation of the moon shadow and its displacement varying with energy \cite{PoS463}. 
Considering the early stage of the experiment, this constitutes a remarkable achievement and demonstrates that the detector is reasonably well understood.

\item IceCube at the South Pole: IceTop, the $1\,$km$^2$ surface array of IceCube, is planned to be enhanced by a hybrid array of scintillators \cite{PoS309} and antennas \cite{PoS401} (additionally small air-Cherenkov telescopes are under investigation \cite{PoS179}).
The coincident detection of air shower particles and radio emission at the surface and high-energy muons in the ice will provide unprecedented accuracy for the reconstruction of air showers.
Therefore, the enhancement can transform IceTop into the most accurate air-shower array of the Southern hemisphere. 
The broad science case extends beyond high-energy Galactic cosmic rays from $10^{14} - 10^{18}\,$eV \cite{PoS332}, and includes the search of PeV photons \cite{PoS184} as well as several improvements for IceCube's neutrino measurements \cite{PoS418}.

\item LOFAR in the Netherlands: LOFAR will get an electronics upgrade increasing the observation time for air showers and an extension of the particle-detector array LORA used as a trigger \cite{PoS363}.
Due to the high density of antennas, LOFAR provides high accuracy for the position of the shower maximum and, thus, for the average mass composition in the energy region of the second knee.

\begin{figure}[t]
    \centering
    \includegraphics[width=0.99\linewidth]{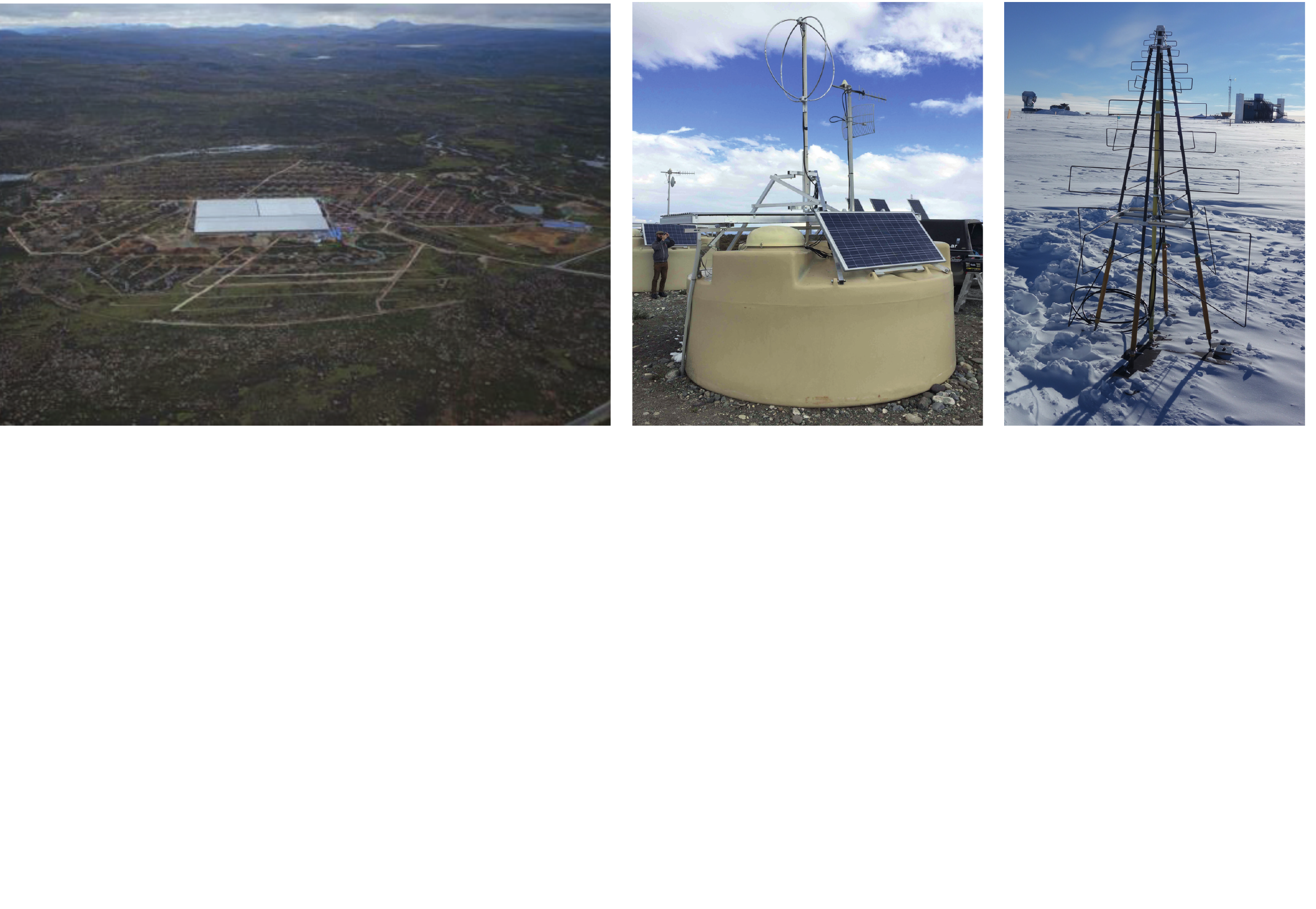}
    \vspace{-5.6cm}
    \caption{Photos of experiments: LHAASO (left, from \cite{PoS693}), a prototype station of AugerPrime (middle, from \cite{PoS395}), antenna of a prototype station for the enhancement of IceTop (right, from \cite{PoS401}). 
}
    \label{fig_Photos}
\end{figure}

\item Space Experiments: Several recent and new space experiments for direct cosmic-ray detections will increase the overlap in energy range with indirect measurements \cite{PoS032}. 
Within the JEM-EUSO program, there are dedicated space experiments planned for air-shower detection at the highest energies using the air-fluorescence and air-Cherenkov techniques \cite{PoS192}. 
In August 2019, Mini-EUSO, a small prototype arrived at the ISS \cite{PoS212}. 
As the next step, a second flight with a super-pressure balloon is planned \cite{PoS247}.
As the final stage, POEMMA is proposed, consisting of two telescopes in nearby satellites, that will exceed the exposure of current ground-based detectors for ultra-high-energy neutrinos and cosmic rays by an order of magnitude \cite{PoS378}. 
Stereo observations will allow for a decent $X_\mathrm{max}$ resolution that enables to measure the average mass composition at the highest energies above $10^{19.5}\,$eV.

\item Telescope Array in Utah, USA: TA will undergo several upgrades extending its energy range. 
TAx4 will quadruple the size of the surface array of scintillation detectors to almost $3000\,$km$^2$ \cite{PoS312}, comparable to the size of Auger. 
Once completed, TA will collect comparable annual exposure at the highest energies.
In particular, anisotropy studies will benefit from this higher statistics for cosmic rays from the Northern sky.
Moreover, TA features two extensions for lower energies: While at high energies the hybrid combination of fluorescence light and scintillation detectors is the method of choice, at lower energies TALE \cite{PoS375} and NICHE \cite{PoS379} use air-Cherenkov light instead of fluorescence light for air-shower observations.
By this, TA will cover five orders of magnitude in energy, reaching from about the knee to the cut-off in the energy spectrum.
Finally, TA remains an excellent site for the test and cross-calibration of new detector systems \cite{PoS259, PoS435, PoS197}.

\item AugerPrime, the ongoing upgrade of the Pierre Auger Observatory in Argentina: 
Next to an extended duty cycle of the fluorescence telescopes, AugerPrime includes various improvements of the $3000\,$km$^2$ surface detector, in particular, the addition of a scintillation detector \cite{PoS434, PoS380} and a radio antenna \cite{PoS274, PoS395} to each water-Cherenkov tank. 
Since the collocated scintillators and water-Cherenkov detectors enable the separation of the electromagnetic and muonic shower components, the upgrade provides Auger with per-event mass sensitivity, even when the fluorescence telescopes are not operating (e.g., during daytime). 
While the scintillators perform well for vertical and mildly inclined events, the antennas in combination with the water-Cherenkov tanks enable a separation of the electromagnetic and muonic shower components for highly inclined events \cite{EPJCradioMuon}. 
Last but not least, underground muon detectors further increase the accuracy in a denser part of the surface array (AMIGA) for the energy range above $10^{17}\,$eV \cite{PoS202}.
Thanks to AugerPrime, Auger will remain the most accurate detector for ultra-high-energy cosmic rays.
The event-by-event mass separation will be used to search for mass-dependent anisotropies.
The higher accuracy for the average mass composition may solve the question to what extent the cut-off in the energy spectrum is due to the GZK effect or due to the maximum acceleration energy of the extragalactic sources.

\item GRAND in China: GRANDproto300 is more than a simple prototype for the huge GRAND array aiming at ultra-high-energy neutrinos \cite{GRANDwhitePaper}. 
GRANDproto300 will target the transition from Galactic-to-extragalactic cosmic rays as its primary science goal \cite{PoS233}. 
The array will be an order of magnitude larger than the AMIGA and AERA arrays of the Pierre Auger Observatory, which enables higher exposure for the energy range around $10^{18}\,$eV.
About 300 antennas shall be deployed over the next two years in China, and are proposed to be complemented by a collocated array of water-Cherenkov detectors. 
Although the sensitivity to vertical showers will be limited because of the sparse antennas spacing, this detector combination of GRANDproto300 will provide mass sensitivity for inclined showers comparable to the radio upgrade of the Pierre Auger Observatory.
\end{itemize}

In summary, most of the upgrades and new experiments rely on hybrid measurements combining two or even more detection techniques (Fig.~\ref{fig_Photos}). 
This trend to multi-hybrid detection is a rational choice, since higher statistics alone may be insufficient for further progress - given the mass composition of cosmic rays turned out to be mixed at all energies up to at least several $10\,$EeV.

\section{Conclusion}
As in the past ICRCs, the CRI session was the one with most contributions. 
Again, they spread over a wide range from the particle physics in showers, over the manifold  instrumentation and their measurements, up to the astrophysical interpretation. 
This reflects a continuous broad and strong interest in this field of astroparticle physics dealing with the highest measured particle energies. 
Indeed, indirect cosmic-ray detection still is the only guaranteed access to these highest energies in the Universe. 
Multi-messenger astronomy is on a rising edge, in particular, the joint interpretation of cosmic rays, photon, and neutrino measurements. 
Nonetheless, at the moment, it is impossible to predict whether the discovery of the most energetic sources in the Milky Way and the Universe will be by one of these neutral messengers or by more accurate cosmic-ray measurements, such as mass-sensitive anisotropy studies, or by a combination of both.
Therefore, it is important to maintain progress in all areas, as is done by the ongoing upgrades and new experiments that will measure air showers with increasingly higher accuracy. 
It is exciting that more and more projects target several messengers with one set of multi-hybrid instrumentation, such as cosmic-ray observatories searching also for neutrinos and photons, or neutrino observatories providing accurate cosmic-ray measurements with their detectors at the same time.
Consequently, it may just be a question of time until the steady progress in our field will pay off in the solution of the puzzle about the origin of cosmic rays.

\small
\subsection*{Acknowledgement}
\noindent
Many thanks to everybody who provided data points and material for my rapporteur talk, and helped me to understand the content presented. 
Further, I truly appreciate the corrections and suggestions for improvements I received by my colleagues and collaborators.
Special thanks to Hans Dembinski, who updated his global spline fit with recent data points (see Fig.~\ref{fig_GSF}), and to the previous rapporteurs helping me with  advice.
Since the majority of contributions falls not directly in my area of expertise, I cannot exclude that I misunderstood some content or mistakenly made wrong interpretations. 
My sincere apologies to all the authors whose valuable contributions I have not mentioned appropriately or left out completely due to space limitations.
Finally, I thank the organizers of the conference and the CRI session for their invitation giving me the opportunity to be a rapporteur at this excellent conference at the marvelous site of the Memorial Union in Madison. 


\begin{thebibliography}{99}

%Energy Spectrum
\bibitem{TibetApj2008}
M. Amenomori et al.~- Tibet AS$\gamma$ Collaboration, \emph{The All-Particle Spectrum of Primary Cosmic Rays in the Wide Energy Range from $10^{14}$ to $10^{17}\,$eV Observed with the Tibet-III Air-Shower Array}, ApJ 678 (2008) 1165.

\bibitem{PoS319}
D. Kostunin for the Tunka-Rex Collaboration, \emph{Seven years of Tunka-Rex operation}, PoS(ICRC2019)319.

\bibitem{Tunka133_7years}
N.M. Budnev et al.~- Tunka-133 Collaboration, \emph{The primary cosmic-ray energy spectrum measured with the Tunka-133 array}, 2019, in preparation.

\bibitem{KG_ICRC2015}
M. Bertaina for the KASCADE-Grande Collaboration, \emph{KASCADE-Grande energy spectrum of cosmic rays interpreted with post-LHC hadronic interaction models}, PoS(ICRC2015)359.

\bibitem{PoS449}
F. Varsi for the GRAPES Collaboration, \emph{Energy spectrum and composition measurements of cosmic rays from GRAPES-3 experiment}, PoS(ICRC2019)449.

\bibitem{HAWC_PRD2017}
R. Alfaro et al.~- HAWC Collaboration, \emph{All-particle cosmic ray energy spectrum measured by the HAWC experiment from 10 to 500 TeV}, PRD 96 (2017) 122001.

\bibitem{PoS318}
R. Koirala for the IceCube Collaboration, \emph{Low Energy Cosmic Ray Spectrum from 250 TeV to 10 PeV using IceTop}, PoS(ICRC2019)318.

\bibitem{PoS172}
K. Andeen for the IceCube Collaboration, \emph{Cosmic Ray Spectrum and Composition from PeV to EeV from the IceCube Neutrino Observatory}, PoS(ICRC2019)172.

\bibitem{PoS298}
D. Ivanov for the Telescope Array Collaboration, \emph{Energy Spectrum Measured by the Telescope Array}, PoS(ICRC2019)298.

\bibitem{PoS450}
V. Verzi for the Pierre Auger Collaboration, \emph{Measurement of the energy spectrum of ultra-high energy cosmic rays using the Pierre Auger Observatory}, PoS(ICRC2019)450.

\bibitem{PoS231}
B. Dawson for the Pierre Auger Collaboration, \emph{The Energy Scale of the Pierre Auger Observatory}, PoS(ICRC2019)231.

\bibitem{TunkaRexLOPESenergyScale}
Tunka-Rex and LOPES Collaborations, \emph{A comparison of the cosmic-ray energy scales of Tunka-133 and KASCADE-Grande via their radio extensions Tunka-Rex and LOPES}, PLB 763(2016)179.

\bibitem{AERAenergyScale}
Pierre Auger Collaboration, \emph{Measurement of the Radiation Energy in the Radio Signal of Extensive Air Showers as a Universal Estimator of Cosmic-Ray Energy}, PRL 116(2016)241101.

\bibitem{PoS234}
O. Deligny for the Pierre Auger and Telescope Array Collaborations, \emph{The energy spectrum of ultra-high energy cosmic rays measured at the Pierre Auger Observatory and at the Telescope Array}, PoS(ICRC2019)234.


\bibitem{PoS176}
J.C. Arteaga-Vel{\'a}zquez for the HAWC Collaboration, \emph{The spectrum of the light component of TeV cosmic rays measured with HAWC}, PoS(ICRC2019)176.

\bibitem{PoS032}
R. Sparvoli, \emph{Cosmic Ray Direct Observations}, PoS(ICRC2019)032.

\bibitem{PoS225}
A. Coleman for the Pierre Auger Collaboration, \emph{Measurement of the Cosmic Ray Flux near the Second Knee with the Pierre Auger Observatory}, PoS(ICRC2019)2335.



% Mass composition
\bibitem{PoS004}
A. Castellina for the Pierre Auger Collaboration, \emph{Highlights from the Pierre Auger Observatory and prospects for AugerPrime}, PoS(ICRC2019)482.

\bibitem{PoS013}
S. Ogio for the Telescope Array Collaboration, \emph{Highlights from the Telescope Array experiment}, PoS(ICRC2019)013.

\bibitem{PoS394}
M. Plum for the IceCube Collaboration, \emph{Cosmic ray composition study using machine learning at the IceCube Neutrino Observatory}, PoS(ICRC2019)394.

\bibitem{PoS306}
D. Kang for the KASCADE-Grande Collaboration, \emph{Latest Results from the KASCADE-Grande Data Analysis}, PoS(ICRC2019)306.


\bibitem{PoS280}
W. Hanlon for the Telescope Array Collaboration, \emph{Telescope Array 10 Year Composition}, PoS(ICRC2019)280.

\bibitem{PoS482}
A. Yushkov for the Pierre Auger Collaboration, \emph{Mass Composition of Cosmic Rays with Energies above $10^{17.2}\,$eV from the Hybrid Data of the Pierre Auger Observatory}, PoS(ICRC2019)482.

\bibitem{PoS440}
J. Todero Peixoto for the Pierre Auger Collaboration, \emph{Estimating the Depth of Shower Maximum using the Surface Detectors of the Pierre Auger Observatory}, PoS(ICRC2019)440.

\bibitem{PoS(ICRC2017)522}
V. de Souza for the Pierre Auger and Telescope Array Collaborations, \emph{Testing the agreement between the $X_\mathrm{max}$ distributions measured by the Pierre Auger and Telescope Array Observatories}, PoS(ICRC2017)522.

\bibitem{UHECR2018_AugerTA_Xmax}
Pierre Auger and Telescope Array Collaborations, \emph{Depth of maximum of air-shower profiles: testing the compatibility of measurements performed at the Pierre Auger Observatory and the Telescope Array experiment}, EPJ WoC 210(2019)01009.


\bibitem{SchroederReview2017}
F.G. Schr{\"o}der, \emph{Radio detection of Cosmic-Ray Air Showers and High-Energy Neutrinos}, PPNP 93(2017)1.

\bibitem{HuegeReview2016}
T. Huege, \emph{Radio detection of cosmic ray air showers in the digital era}, Phys. Rept. 620(2016)1.

\bibitem{PoS205}
S. Buitink et al. - LOFAR, \emph{Towards an improved mass composition analysis with LOFAR}, PoS(ICRC2019)205. 

\bibitem{PoS246}
A. Escudie for the CODALEMA Collaboration, \emph{From the Observation of UHECR Radio Signal in [1-200] MHz to the Composition: CODALEMA and EXTASIS Status Report}, PoS(ICRC2019)246.

\bibitem{PoS362}
K. Mulrey et al. - LOFAR, \emph{The energy scale of cosmic rays detected with LOFAR}, PoS(ICRC2019)362.

\bibitem{PoS331}
V. Lenok for the Tunka-Rex Collaboration, \emph{Modeling the Aperture of Radio Instruments for Air-Shower Detection}, PoS(ICRC2019)331.



\bibitem{GSF_ICRC2017}
H. Dembinski et al., \emph{Data-driven model of the cosmic-ray flux and mass composition from $10\,$GeV to $10^{11}\,$GeV}, PoS(ICRC2017)533.



% Anisotropy

\bibitem{PoS014}
D. Soldin for the IceCube Collaboration, \emph{Recent Results of Cosmic Ray Measurements from IceCube and IceTop}, PoS(ICRC2019)014.

\bibitem{HAWC_IceCube_anisotropy}
IceCube and HAWC Collaborations, \emph{All-sky Measurement of the Anisotropy of Cosmic Rays at 10 TeV and Mapping of the Local Interstellar Magnetic Field}, APJ 871(2019)96.

\bibitem{AhlersAnisotropyReview2017},
M. Ahlers et al., \emph{Origin of small-scale anisotropies in Galactic cosmic rays}, PPNP 94(2017)184.

\bibitem{AugerScience2017}
Pierre Auger Collaboration, \emph{Observation of a large-scale anisotropy in the arrival directions of cosmic rays above 8 $\times$ 10$^{18}$ eV}, Science 357(2017)1266.

\bibitem{PoS488}
Y. Zhang for the Tibet AS$\gamma$ Collaboration, \emph{Large-scale Cosmic Ray Anisotropy with Tibet air shower array}, PoS(ICRC2019)488.

\bibitem{PoS263}
W. Gao for the LHAASO Collaboration, \emph{The Large-scale Anisotropy of Cosmic Rays Observed with the Partial LHAASO-KM2A Array}, PoS(ICRC2019)263.


\bibitem{PoS408}
E. Roulet for the Pierre Auger Collaboration, \emph{Large-scale anisotropies above 0.03 EeV measured by the Pierre Auger Observatory}, PoS(ICRC2019)408.

\bibitem{PoS206}
L. Caccianiga for the Pierre Auger Collaboration, \emph{Anisotropies of the Highest Energy Cosmic-ray Events Recorded by the Pierre Auger Observatory in 15 years of Operation}, PoS(ICRC2019)206.


\bibitem{PoS468}
T. Winchen et al., \emph{Modification of the Energy Spectrum of UHECR by the Galactic Magnetic Field for Anisotropic Arrival Directions}, PoS(ICRC2019)468.

\bibitem{PoS310}
K. Kawata for the Telescope Array Collaboration, \emph{Updated Results on the UHECR Hotspot Observed by the Telescope Array Experiment}, PoS(ICRC2019)310.

\bibitem{PoS439}
A. di Matteo for the Pierre Auger and Telescope Array Collaborations, \emph{Full-sky searches for anisotropies in UHECR arrival directions with the Pierre Auger Observatory and the Telescope Array}, PoS(ICRC2019)439.






% Origin
\bibitem{PoS209}
D. Caprioli and C. Haggerty, \emph{The Issue with Diffusive Shock Acceleration}, PoS(ICRC2019)209.

\bibitem{PoS228}
C. Cotter et al., \emph{Acceleration of He nuclei at non-relativistic collisionless shocks}, PoS(ICRC2019)228.

\bibitem{PoS315}
J. Kim et al., \emph{Propagation of Ultra-high-energy Cosmic Rays in the Magnetized Cosmic Web}, PoS(ICRC2019)315.

\bibitem{PoS364}
M.S. Muzio et al., \emph{Constraints on UHECR sources and their environments, from fitting UHECR spectrum and composition, and neutrinos and gammas}, PoS(ICRC2019)364.

\bibitem{PoS196}
D. Biehl et al., \emph{Gamma-Ray Bursts as Sources of Ultra-High Energy Cosmic Rays across the Ankle}, PoS(ICRC2019)196.

\bibitem{UngerPRD2015}
M. Unger et al., \emph{Origin of the ankle in the ultrahigh energy cosmic ray spectrum, and of the extragalactic protons below it}, PRD 92(2015)123001.


\bibitem{PoS031}
J. Sitarek et al., \emph{Gamma-ray Indirect Rapporteur}, PoS(ICRC2019)031.

\bibitem{HESS_GC_Nature2016}
H.E.S.S.~Collaboration, \emph{Acceleration of petaelectronvolt protons in the Galactic Centre}, Nature 531(2016)476.



% Hadronic Models
\bibitem{PoS235}
H. Dembinski et al., \emph{Future Proton-Oxygen Beam Collisions at the LHC for Air Shower Physics}, PoS(ICRC2019)235.

\bibitem{PoS005}
L. Cazon, \emph{Probing High-Energy Hadronic Interactions with Extensive Air Showers}, PoS(ICRC2019)005.

\bibitem{PoS214}
L. Cazon for the EAS-MSU, IceCube, KASCADE-Grande, NEVOD-DECOR, Pierre Auger, SUGAR, Telescope Array, and Yakutsk EAS Array Collaborations, \emph{Working Group Report on the Combined Analysis of Muon Density Measurements from Eight Air Shower Experiments}, PoS(ICRC2019)214.

\bibitem{PoS411}
F.A. S{\'a}nchez for the Pierre Auger Collaboration, \emph{The muon component of extensive air showers above $10^{17.5}\,$eV measured with the Pierre Auger Observatory}, PoS(ICRC2019)411.

\bibitem{PoS404}
F. Riehn for the Pierre Auger Collaboration, \emph{Measurement of the fluctuations in the number of muons in inclined air showers with the Pierre Auger Observatory}, PoS(ICRC2019)404.

\bibitem{PoS177}
J.C. Arteaga-Vel{\'a}zquez for the KASCADE-Grande Collaboration, \emph{Muon content in air showers between 10 PeV and 1 EeV determined from measurements with KASCADE-Grande}, PoS(ICRC2019)177.

\bibitem{PoS351}
A. Mitchell et al., \emph{Using IACTs to Measure the Profiles of Muons in TeV Air Showers}, PoS(ICRC2019)351.

\bibitem{PoS417}
H. Schoorlemmer et al., \emph{Differences between High Energy Hadronic Interaction Models for Air Shower Measurements in the 100 GeV-100 TeV Range}, PoS(ICRC2019)417.


\bibitem{PoS446}
M. Unger for the NA61/SHINE Collaboration, \emph{New Results from the Cosmic-Ray Program of the NA61/SHINE facility at the CERN SPS}, PoS(ICRC2019)446. 

\bibitem{PoS207}
F. Cafagna for the TOTEM Collaboration, \emph{Latest results for Proton-proton Cross Section Measurements with the TOTEM experiment at LHC}, PoS(ICRC2019)207.

\bibitem{PoS349}
H. Menjo et al., \emph{The results and future prospects of the LHCf experiment}, PoS(ICRC2019)349. 

\bibitem{PoS188}
S. Baur for the CMS Collaboration, \emph{Measurements of the very-forward energy in pp collisions at the LHC and constraints for cosmic ray air showers}, PoS(ICRC2019)188. 

\bibitem{PoS387}
T. Pierog et al., \emph{Collective Hadronization and Air Showers: Can LHC Data Solve the Muon Puzzle?}, PoS(ICRC2019)387. 


\bibitem{PoS894}
T. Gaisser for the IceCube Collaboration, \emph{Seasonal Variation of Atmospheric Muons in IceCube}, PoS(ICRC2019)894.

\bibitem{PoS226}
L. Cazon et al., \emph{Probing the High Energy Spectrum of Neutral Pions in Ultra-high-energy Proton-Air Interactions}, PoS(ICRC2019)226.

\bibitem{PoS236}
H. Dembinski for the CORSIKA 8 Collaboration, \emph{Technical Foundations of CORSIKA 8: New Concepts for Scientific Computing}, PoS(ICRC2019)236.

\bibitem{PoS181}
D. Baack for the CORSIKA 8 Collaboration, \emph{Performance optimization of air shower simulations with CORSIKA}, PoS(ICRC2019)181.

\bibitem{PoS399}
D. Melo for the CORSIKA 8 Collaboration, \emph{First results of the CORSIKA 8 air shower simulation framework}, PoS(ICRC2019)399.




% Other
\bibitem{PoS377}
L. Olah et al., \emph{Improvement of cosmic-ray muography for Earth sciences and civil engineering}, PoS(ICRC2019)377.

\bibitem{PoS381}
J. Pe{\~n}a Rodr{\'i}guez et al., \emph{Calibration and first measurements of MuTe: a hybrid Muon Telescope for geological structures}, PoS(ICRC2019)381.

\bibitem{PoS275}
V. Grabsky et al., \emph{Prototype-module of a muon tracker to investigate the density distribution of the Popocatepetl volcano lava dome}, PoS(ICRC2019)275.

\bibitem{PoS201}
G. Bonomi et al., \emph{A Cosmic Rays Tracking System for the Stability Monitoring of Historical Buildings}, PoS(ICRC2019)201.

\bibitem{PoS416}
O. Scholten et al., \emph{Influence of atmospheric electric fields on radio emission from air showers}, PoS(ICRC2019)416.

\bibitem{PoS018}
R. Abbasi, \emph{Cosmic Ray Detectors and Observational Breakthroughs in Atmospheric Electricity}, PoS(ICRC2019)018.


\bibitem{PoS428}
K. Smelcerz for the CREDO Collaboration, \emph{A communication solution for portable detectors of the Cosmic Ray Extremely Distributed Observatory}, PoS(ICRC2019)428.


\bibitem{PoS284}
A. Haungs et al., \emph{German-Russian Astroparticle Data Life Cycle Initiative}, PoS(ICRC2019)284.

\bibitem{KCDC_EPJC}
KASCADE-Grande Collaboration. \emph{The KASCADE Cosmic-ray Data Centre KCDC: granting open access to astroparticle physics research data}, EPJ C 78(2018)741.

\bibitem{PoS453}
J. Vicha et al., \emph{Invisible Energy from KASCADE Data}, PoS(ICRC2019)453.






% Future Detectors
\bibitem{Astro2020_GCR}
F.G. Schroeder et al., \emph{High-Energy Galactic Cosmic Galactic Cosmic Rays}, BAAS 51(3) (2019) 131, arXiv:1903.07713.

\bibitem{Astro2020_UHECR}
F. Sarazin et al., \emph{What is the origin of the highest-energy particles in the universe?}, BAAS 51(3) (2019) 093, arXiv:1903.04063.


\bibitem{PoS217}
M. Chen for the LHAASO Collaboration, \emph{Status and First Result of LHAASO-WCDA}, PoS(ICRC2019)217.

\bibitem{PoS219}
S. Chen for the LHAASO Collaboration, \emph{Detector Simulation of LHAASO-KM2A with Gean}, PoS(ICRC2019)219.

\bibitem{PoS472}
G. Xin for the LHAASO Collaboration, \emph{Study on the muon lateral distribution based on the first stage of LHAASO-KM2A}, PoS(ICRC2019)472.

\bibitem{PoS463}
Y. Wang for the LHAASO Collaboration, \emph{The Energy Calibration Using the Moon Shadow of LHAASO-WCDA Detector}, PoS(ICRC2019)463.

\bibitem{PoS693}
H. He for the LHAASO Collaboration, \emph{Status and First Results of the LHAASO Experiment}, PoS(ICRC2019)693.



\bibitem{PoS309}
M. Kauer for the IceCube Collaboration, \emph{The Scintillator Upgrade of IceTop: Performance of the Prototype Array}, PoS(ICRC2019)309.

\bibitem{PoS401}
M. Renschler for the IceCube Collaboration, \emph{First measurements with prototype radio antennas for the IceTop detector array}, PoS(ICRC2019)401.

\bibitem{PoS179}
M. Schaufel for the IceCube Collaboration, \emph{IceAct, small Imaging Air Cherenkov Telescopes for IceCube}, PoS(ICRC2019)179.

\bibitem{PoS332}
A. Leszczynska for the IceCube Collaboration, \emph{Simulation and Reconstruction Study of a Future Surface Scintillator Array at the IceCube Neutrino Observatory}, PoS(ICRC2019)332.

\bibitem{PoS184}
A. Balagopal V. et al., \emph{Frequency-optimised radio air arrays for air-shower detection}, PoS(ICRC2019)184. 

\bibitem{PoS418}
F. Schroeder for the IceCube Collaboration, \emph{Science Case of a Scintillator and Radio Surface Array at IceCube}, PoS(ICRC2019)418.


\bibitem{PoS363}
K. Mulrey et al. - LOFAR, \emph{Extension of the LOFAR Radboud Air Shower Array}, PoS(ICRC2019)363. 


\bibitem{PoS192}
M.E. Bertaina for the JEM-EUSO Collaboration, \emph{Search for Ultra-High Energy Cosmic Rays from Space - the JEM-EUSO Program}, PoS(ICRC2019)192. 

\bibitem{PoS212}
M. Casolino for the JEM-EUSO Collaboration, \emph{Mini-EUSO experiment to study UV emission of terrestrial and astrophysical origin onboard of the International Space Station}, PoS(ICRC2019)212. 

\bibitem{PoS247}
J. Eser for the JEM-EUSO Collaboration, \emph{Results of the EUSO-SPB1 flight}, PoS(ICRC2019)247. 

\bibitem{PoS378}
A.V. Olinto et al., \emph{POEMMA: Probe Of Extreme Multi-Messenger Astrophysics}, PoS(ICRC2019)378.


\bibitem{PoS312}
E. Kido for the Telescope Array Collaboration, \emph{Status and prospects of the TAx4 experiment}, PoS(ICRC2019)312.

\bibitem{PoS375}
S. Ogio for the Telescope Array Collaboration, \emph{Telescope Array Low energy Extension(TALE) Hybrid}, PoS(ICRC2019)375.

\bibitem{PoS379}
Y. Omur for the Telescope Array Collaboration, \emph{NICHE detector and operations}, PoS(ICRC2019)379.


\bibitem{PoS259}
T. Fujii for the FAST Collaboration, \emph{Observing ultra-high energy cosmic rays with prototypes of Fluorescence detector Array of Single-pixel Telescopes (FAST) in both hemispheres}, PoS(ICRC2019)259.

\bibitem{PoS435}
Y. Tameda et al., \emph{The status and performance of Cosmic Ray Air Fluorescence Fresnel lens Telescope (CRAFFT) for the next generation UHECR observatory}, PoS(ICRC2019)435.

\bibitem{PoS197}
F. Bisconti for the JEM-EUSO Collaboration, \emph{EUSO-TA ground based fluorescence detector: analysis of the detected events}, PoS(ICRC2019)192. 




\bibitem{PoS434}
A. Taboada for the Pierre Auger Collaboration, \emph{Analysis of Data from Surface Detector Stations of the AugerPrime Upgrade}, PoS(ICRC2019)434.

\bibitem{PoS380}
J. Pekala for the Pierre Auger Collaboration, \emph{Production and Quality Control of the Scintillator Surface Detector for the AugerPrime Upgrade of the Pierre Auger Observatory}, PoS(ICRC2019)380.

\bibitem{PoS274}
M. Gottowik for the Pierre Auger Collaboration, \emph{Measurements of Inclined Air Showers with the Auger Engineering Radio Array at the Pierre Auger Observatory}, PoS(ICRC2019)274.

\bibitem{PoS395}
B. Pont for the Pierre Auger Collaboration, \emph{A Large Radio Detector at the Pierre Auger Observatory - Measuring the Properties of Cosmic Rays up to the Highest Energies}, PoS(ICRC2019)395.

\bibitem{EPJCradioMuon}
E.M. Holt et al., \emph{Enhancing the cosmic-ray mass sensitivity of air-shower arrays by combining radio and muon detector}, EPJ C 79(2019)371.

\bibitem{PoS202}
A.M. Botti for the Pierre Auger Collaboration, \emph{The AMIGA underground muon detector of the Pierre Auger Observatory - performance and event reconstruction}, PoS(ICRC2019)202.



\bibitem{GRANDwhitePaper}
GRAND Collaboration, \emph{The Giant Radio Array for Neutrino Detection (GRAND): Science and Design}, Sci.~China Phys.~Mech.~Astron.~63(2020)219501.

\bibitem{PoS233}
V. Decoene for the GRAND Collaboration, \emph{The GRANDProto300 experiment: a pathfinder with rich astroparticle and radio-astronomy science case}, PoS(ICRC2019)233. 







\end{thebibliography}
\end{document}